\documentclass[english,aps,prX,showpacs,twocolumn]{revtex4-1}
\usepackage[T1]{fontenc}
\usepackage[latin1]{inputenc}
\usepackage{tikz}
\usepackage{verbatim}
\usepackage{amsmath}
\usepackage[english]{babel}
\usepackage{bm}
\usepackage{bbm}
\usepackage{tikz}
\usepackage{amssymb}
\usepackage{pgfplots}
\usepackage{verbatim}
\usetikzlibrary{calc}
\usepackage[sort&compress]{natbib}
\usepackage{lipsum}
\usepackage{graphicx}
\usetikzlibrary{arrows}
\usetikzlibrary{calc}
\usepackage{ifthen}
\usetikzlibrary{shapes.geometric,positioning,patterns,calc}
\usetikzlibrary{decorations.markings}

\usepackage{color}

\pgfmathdeclarefunction{gauss}{2}{%
  \pgfmathparse{1/(#2*sqrt(2*pi))*exp(-((x-#1)^2)/(2*#2^2))}%
}

\newcommand{\boundellipse}[3] 
{(#1) ellipse (#2 and #3)
}

\begin{document}

\begin{abstract}
The influence of possible magnetic inertia effects has recently drawn attention in ultrafast magnetization dynamics and switching. Here we derive rigorously a description of inertia in the Landau-Lifshitz-Gilbert equation on the basis of the Dirac-Kohn-Sham framework. Using the Foldy-Wouthuysen transformation up to the order of $1/c^4$ gives the intrinsic inertia of a pure system
through the 2$^{\rm nd}$ order time-derivative of magnetization in the dynamical equation of motion. Thus, the inertial damping $\mathcal{I}$ is a higher order spin-orbit coupling effect, $\sim 1/c^4$, as compared to the Gilbert damping $\Gamma$ that is of order $1/c^2$. 
Inertia is therefore expected to play a role only on ultrashort timescales (sub-picoseconds). We also show that the Gilbert damping and inertial damping are related to one another through the imaginary and real parts of the magnetic susceptibility tensor respectively.

\end{abstract}

\title{Relativistic theory of magnetic inertia in ultrafast spin dynamics}

\author{Ritwik Mondal}
\email[]{ritwik.mondal@physics.uu.se}
\author{Marco Berritta}
\author{Ashis K. Nandy}
\author{Peter M. Oppeneer}
\affiliation{Department of Physics and Astronomy, Uppsala University, P.\,O.\ Box 516,  SE-75120 Uppsala, Sweden}

\pacs{71.15.Rf, 75.78.-n, 75.40.Gb}
\date{\today} 
 
\maketitle
\section{Introduction}

 The foundation of contemporary magnetization dynamics is the Landau-Lifshitz-Gilbert (LLG) equation which describes the precession of spin moment and a transverse damping of it, while keeping the modulus of magnetization vector fixed \cite{landau35,gilbert56,gilbert04}. The LLG equation of motion was originally derived phenomenologically and the damping of spin motion has been attributed to relativistic effects such as the spin-orbit interaction \cite{landau35,kambersky70,kambersky76,kunes02}. In recent years  there has been a flood of proposals for the fundamental microscopic mechanism behind the Gilbert damping: the breathing Fermi surface model of Kambersk\'y, where the damping is due to magnetization precession and the effect of spin-orbit interaction at the Fermi surface \cite{kambersky70}, the extension of the breathing Fermi surface model to the  torque-torque correlation model \cite{kambersky76,kambersky07}, scattering theory description \cite{Brataas2008}, effective field theories \cite{Fahnle2011JPCM}, linear response formalism within relativistic electronic structure theory \cite{Ebert2011}, and the Dirac Hamiltonian theory formulation \cite{Mondal2016}.

For practical reasons it was needed to extend the original LLG equation to include several other mechanisms \cite{Berger1996,zhang09}. To describe e.g.\ current induced spin-transfer torques, the effects of spin currents have been taken into account \citep{zhang04,Tatara2004,Duine2007}, as well as spin-orbit torques \cite{Freimuth2015}, and the effect of spin diffusion \cite{Vignale2009}. A different kind of spin relaxation due to the exchange field has been introduced by Bar'yakhtar {\it et al.}\ \cite{baryakhtar84}. In the Landau-Lifshitz-Bar'yakhtar equation spin dissipations originate from the spatial dispersion of exchange effects through the second order space derivative of the effective field \cite{baryakhtar13r,baryakhtarultrafast13}. A further recent work predicts the existence of extension terms that contain spatial as well temporal derivatives of the local magnetization \cite{Li2016}. 

Another term, not discussed in the above investigations, is the magnetic inertial damping that has recently drawn attention \cite{Kimel2009,Ciornei2011,Ciornei2010thesis}. 
Originally, magnetic inertia was discussed following the discovery of earth's magnetism \cite{Walker1917}. 
Within the LLG framework, inertia is introduced as an additional term \cite{Ciornei2011,faehnle11,Wegrowe2012,Wegrowe2015JAP} 
leading to a modified LLG equation, 
\begin{equation}
\frac{\partial \bm{M}}{\partial t} =- \gamma  \bm{M} \times \bm{H}_{\rm eff}   +   \bm{M} \times \left(\Gamma \frac{\partial \bm{M}}{\partial t } + \mathcal{I}\,\frac{\partial^2 \bm{M}}{\partial^2 t} \right),
\label{LLG}
\end{equation}
where $\Gamma$ is the 
Gilbert damping constant \cite{landau35,gilbert56,gilbert04}, $\gamma$ the gyromagnetic ratio, $\bm{H}_{\rm eff}$ the effective magnetic field, and
 $\mathcal{I}$ is the inertia of the magnetization dynamics, similar to the mass in Newton's equation. This type of motion has  the same classical analogue as the  nutation of a spinning symmetric top. 
The potential importance of inertia is illustrated in Fig.\ 1. While Gilbert damping slowly aligns the precessing magnetization to the effective magnetic field, inertial dynamics causes a trembling or nutation of the magnetization vector \cite{Henk2012,Ciornei2011,Bhattacharjee2012}. 
Nutation could consequently pull the magnetization toward the equator and cause its switching to the antiparallel direction \cite{Tatara2015,thonig2016}, whilst depending crucially on the strength of the magnetic inertia. 
The parameter $\mathcal{I}$ that characterizes the nutation motion is in the most general case a tensor and has been associated 
with the magnetic susceptibility \cite{Bhattacharjee2012,Wegrowe2015JAP,thonig2016}. 
Along a different line of reasoning, F\"ahnle {\it et al.}\  extended the breathing Fermi surface model to include the effect of magnetic inertia \cite{faehnle11,fahnle2013PRB}.
The technological importance of nutation dynamics is thus its potential to steer magnetization switching in memory devices \cite{Kimel2009,Ciornei2011,Ciornei2010thesis,Tatara2015} and also in skyrmionic spin textures \cite{Buttner2015}. Magnetization dynamics involving inertial dynamics has been investigated recently and it was suggested that its dynamics belongs to smaller time-scales i.e., the femtosecond regime \cite{Ciornei2011}. However, the origin of inertial damping from a fundamental framework is still missing, and, moreover, although it is possible to vary the size of the inertia in spin-dynamics simulations, it is unknown what the typical size of the inertial damping is.

Naturally the question arises whether it is possible to derive the extended LLG equation including inertia while starting from the fully relativistic Dirac equation. Hickey and Moodera \cite{Hickey2009} started from a Dirac Hamiltonian and obtained an intrinsic Gilbert damping term which originated from spin-orbit coupling. 
However they started from only a part of the spin-orbit coupling Hamiltonian which was anti-hermitian \cite{widom2009,Hickey_reply2009}. 
A recent derivation based on Dirac Hamiltonian theory formulation \cite{Mondal2016} showed that the Gilbert damping depends strongly on both interband and intraband transitions (consistent with Ref.\ \cite{gilmore07}) as well as the magnetic susceptibility response function, $\chi_m$.
This derivation used the relativistic expansion to the lowest order $1/c^2$ of the hermitian Dirac-Kohn-Sham (DKS) Hamiltonian including the effect of exchange field \cite{Mondal2015a}. 

\begin{figure}[t!]
\centering 
\includegraphics[width = 0.47\textwidth]{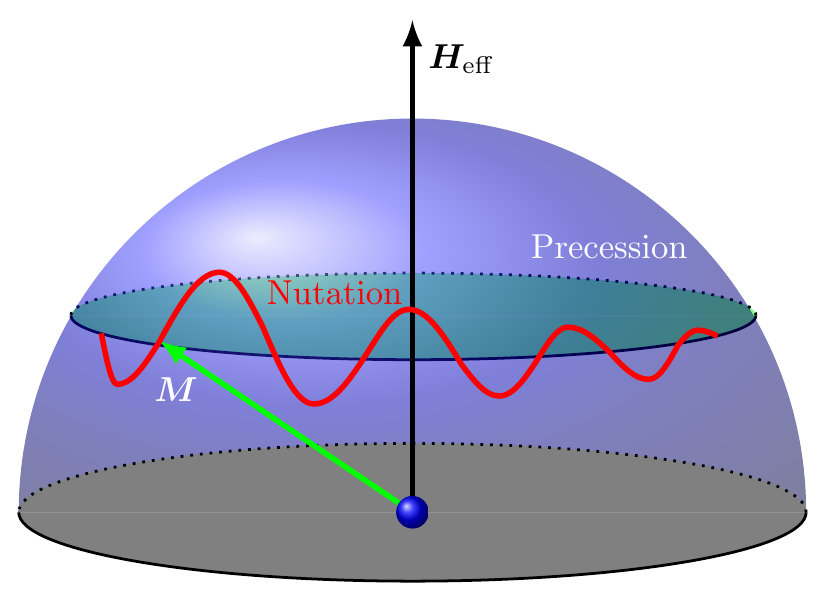}
\caption{(Color online) Schematic illustration of magnetization dynamics. The precessional motion of $\bm{M}$ around $\bm{H}_{\rm eff}$ is depicted by the blue solid-dashed curve and the nutation is shown by the red curve.}
\end{figure}

In this article we follow an approach similar to that of Ref.\ \cite{Mondal2016} but we consider higher order expansion terms of the DKS Hamiltonian up to the order of $1/c^4$. This is shown to lead to the intrinsic inertia term in the modified LLG equation and demonstrates that it stems from a higher-order spin-orbit coupling term.
A relativistic origin of the spin nutation angle, caused by Rashba-like spin-orbit coupling, was  previously concluded, too, in the context of semiconductor nanostructures 
 \cite{Cho2008,Winkler2004}.

In the following, we derive in Sec.\ \ref{1} the relativistic correction terms to the extended Pauli Hamiltonian up to the order of $1/c^4$, which includes the spin-orbit interaction and an additional term. Then the corresponding magnetization dynamics is computed from the obtained spin Hamiltonian in Sec.\ \ref{2}, which is shown to contain the Gilbert damping and the magnetic inertial damping. Finally, we discuss the size of the magnetic inertia in relation to other earlier studies.

\section{Relativistic Hamiltonian formulation}
\label{1}
We start our derivation with a fully relativistic particle, a Dirac particle \cite{Dirac1928} inside a material and in the presence of an external field, for which we write the DKS Hamiltonian:
\begin{eqnarray}
\label{Dirac_Hamiltonian}
\mathcal{H}&=& c\underline{\bm{\alpha}}\cdot\left(\bm{p}-e\bm{A}\right)+(\underline{\beta}-\underline{\mathbbm{1}}) mc^{2}+ V \underline{\mathbbm{1}} \nonumber\\
& =& \mathcal{O}+(\underline{\beta}-\underline{\mathbbm{1}}) mc^{2}+\mathcal{E} ,
\end{eqnarray}
where $V$ is the effective crystal potential created by the ion-ion, ion-electron and electron-electron interactions, $\bm{A}(\bm{r},t)$ is the magnetic vector potential from the external field, $c$ is the speed of light, $m$ is particle's mass and $\underline{\mathbbm{1}}$ is the $4\times 4$ unit matrix.  
 $\underline{\bm{\alpha}}$ and $\underline{\beta}$  are the Dirac matrices that have the form
\[
\underline{\bm{\alpha}}=\left(\begin{array}{cc}
0 & \bm{\sigma}\\
\bm{\sigma} & 0
\end{array}\right),\quad\mbox{}\quad\underline{\beta}=\left(\begin{array}{cc}
\bm{1} & 0\\
0 & -\bm{1}
\end{array}\right)\,,
\]
where $\bm{\sigma}$ is the Pauli spin matrix vector and $\bm{1}$ is $2\times 2$ unit matrix. The Dirac equation is then written as $i\hbar\frac{\partial \psi(\bm{r},t)}{\partial t} = \mathcal{H}\psi(\bm{r},t)$ for a Dirac bi-spinor $\psi$. The quantity 
$\mathcal{O}=c\underline{\bm{\alpha}}\cdot\left(\bm{p}-e\bm{A}\right)$ defines the off-diagonal, or odd terms in the matrix formalism  and  $\mathcal{E}=V \underline{\mathbbm{1}}$ are the diagonal, i.e., even terms. The latter have to be multiplied by a $2\times 2$ block diagonal unit matrix in order to bring them in a matrix form. To obtain the nonrelativistic Hamiltonian and the relativistic corrections 
one can write down the Dirac bi-spinor in double two component form as 
\[\psi(\bm{r},t) = \left(\begin{array}{cc}
\phi(\bm{r},t)\\
\eta(\bm{r},t) 
\end{array} \right), \]
and substitute those into the Dirac equation. The upper two components represent the particle and the lower two components represent the anti-particle. 
However the question of separating the particle's and anti-particle's wave functions is not clear for any given momentum. As the part $\underline{\bm{\alpha}}\cdot\bm{p}$ is off-diagonal in the matrix formalism, it retains the odd components and thus links the particle-antiparticle wave function. One way to eliminate the antiparticle's wave function is by an exact transformation \cite{kraft95} which gives terms that require a further expansion in powers of $1/c^2$.  Another way is to search for a representation where the odd terms become smaller and smaller and one can ignore those with respect to the even terms and retain only the latter \cite{Strange1998}. 
The Foldy-Wouthuysen (FW) transformation \cite{Foldy1950,greiner2000relativistic} was the very successful attempt to find such a representation.

It is an unitary transformation obtained by suitably choosing the FW operator, 
\begin{eqnarray}
\label{unitary_operator}
U_{\rm FW} &=& -\frac{i}{2mc^2}\underline{\beta}\mathcal{O}.
\end{eqnarray}
The minus sign in front of the operator is because $\underline{\beta}$ and $\mathcal{O}$ anti-commute with each other.
The transformation of the wave function adopts the form $\psi^{\prime}(\bm{r},t) = e^{iU_{\rm FW}}\psi(\bm{r},t)$ such that the probability density remains the same, $\vert \psi\vert^2 = \vert\psi^{\prime} \vert^2$. The time-dependent FW transformation can be expressed as \cite{Strange1998,Silenko2016PRA}
\begin{eqnarray}
\label{FW_transformation}
\mathcal{H}_{\rm FW}&=& e^{iU_{\rm FW}}\left(\mathcal{H}-i\hbar\frac{\partial}{\partial t}\right)e^{-iU_{\rm FW}} + i\hbar \frac{\partial}{\partial t}\,.
\end{eqnarray}
The first term can be expanded in a series as
\begin{eqnarray}
&e^{iU_{\rm FW}}\mathcal{H} e^{-iU_{\rm FW}} = \mathcal{H}+i\left[U_{\rm FW},\mathcal{H}\right]+\frac{i^2}{2!}\left[U_{\rm FW},\left[U_{\rm FW},\mathcal{H}\right]\right]\nonumber\\
&+....+ \frac{i^n}{n!}\left[U_{\rm FW},\left[U_{\rm FW}, ... \left[U_{\rm FW},\mathcal{H}\right]...\right]\right]+...\,\,.
\end{eqnarray}
The time dependency enters through the second term of Eq.\  (\ref{FW_transformation}) and for a time-independent transformation one works with $\frac{\partial U_{\rm FW}}{\partial t} = 0$. It is instructive to note that the aim of the whole procedure is to make the odd terms smaller and one can notice that as it goes higher and higher in the expansion, the corresponding coefficients decrease of the order $1/c^2$ due to the choice of the unitary operator. After a first transformation, the new Hamiltonian will contain new even terms, $\mathcal{E}^\prime$, as well as new odd terms, $\mathcal{O}^\prime$ of $1/c^2$ or higher. The latter terms can be used to perform a next transformation having the new unitary operator as $U^{\prime}_{\rm FW} = -\frac{i}{2mc^2}\underline{\beta}\mathcal{O}^{\prime}$. After a second transformation the new Hamiltonian, $\mathcal{H}^{\prime}_{\rm FW}$ is achieved that has the odd terms of the order $1/c^4$ or higher. The transformation is a repetitive process and it continues until the separation of positive and negative energy states are guaranteed.     

After a fourth transformation we derive the new transformed Hamiltonian with all the even terms that are correct up to the order of $\frac{1}{m^3c^6}$ as
\cite{schwabl2008advanced,Silenko2013,Silenko2016PRA}
\begin{widetext}
\begin{eqnarray}
\mathcal{H}^{\prime\prime\prime}_{\rm FW} &=&(\underline{\beta} -\underline{\mathbbm{1}})mc^2 + \underline{\beta}\left(\frac{\mathcal{O}^2}{2mc^2}-\frac{\mathcal{O}^4}{8m^3c^6}\right) +  \mathcal{E} - \frac{1}{8m^2c^4}\left[\mathcal{O},\left[\mathcal{O},\mathcal{E}\right]+i\hbar\,\dot{\mathcal{O}}\right]\nonumber\\
& + & \frac{\underline{\beta}}{16m^3c^6}\left\{\mathcal{O},\left[\left[\mathcal{O},\mathcal{E}\right],\mathcal{E}\right]\right\} + \frac{\underline{\beta}}{8m^3c^6}\left\{\mathcal{O},\left[i\hbar\,\dot{\mathcal{O}},\mathcal{E}\right]\right\} + \frac{\underline{\beta}}{16m^3c^6}\left\{\mathcal{O},\left(i\hbar\right)^2\,\ddot{\mathcal{O}}\right\} .
\label{FW-Hamil}
\end{eqnarray}
Note that $[A,B]$ defines the commutator, while $\{A,B\}$ represents the anti-commutator for any two operators $A$ and $B$. A similar Foldy-Wouthuysen transformation Hamiltonian up to an order of $1/m^3 c^6$ was derived by Hinschberger and Hervieux in their recent work \cite{Hinschberger2012}, however there are some differences, for example, the first and second terms in the second line of Eq.\  (\ref{FW-Hamil}) were not given. 
Once we have the transformed Hamiltonian as a function of odd and even terms, the final form is achieved by substituting the correct form of odd terms $\mathcal{O}$ and calculating term by term. 

Evaluating all the terms separately, we derive the Hamiltonian for only the positive energy solutions i.e. the upper components of the Dirac bi-spinor as a $2\times 2$ matrix formalism \cite{Mondal2015a,Hinschberger2012,hinschberger13}:
\begin{eqnarray}
\label{final_FW}
\mathcal{H}^{\prime\prime\prime}_{\rm FW} & = & \frac{\left(\bm{p}-e\bm{A}\right)^{2}}{2m}+V-\frac{e\hbar}{2m}\bm{\sigma}\cdot\bm{B}-\frac{\left(\bm{p}-e\bm{A}\right)^{4}}{8m^{3}c^{2}}
-\frac{e\hbar^{2}}{8m^{2}c^{2}}\bm{\nabla}\cdot\bm{E}_{\rm tot} + \frac{e\hbar}{8m^3c^2}\left\{\left(\bm{p}-e\bm{A}\right)^2,\bm{\sigma}\cdot \bm{B} \right\}\nonumber\\
&-& \frac{e\hbar}{8m^{2}c^{2}}\bm{\sigma}\cdot\left[ \bm{E}_{\rm tot}\times\left(\bm{p}-e\bm{A}\right)-\left(\bm{p}-e\bm{A}\right)\times\bm{E}_{\rm tot}\right] \nonumber\\
&-& \frac{e\hbar^2}{16m^3c^4} \left\{\left(\bm{p}-e\bm{A}\right),\partial_t\bm{E}_{\rm tot} \right\}
- \frac{ie\hbar^2}{16m^3c^4}\bm{\sigma}\cdot\left[ \partial_t\bm{E}_{\rm tot}\times\left(\bm{p}-e\bm{A}\right)+\left(\bm{p}-e\bm{A}\right)\times\partial_t\bm{E}_{\rm tot}\right] ,\,\,\,
\end{eqnarray}
\end{widetext}
where $\partial_t \equiv \partial/\partial t$ defines the first-order time derivative. 
 The higher order terms ($1/c^6$ or more) will involve similar formulations and more and more time derivatives of the magnetic and electric fields will appear that stem from the time derivative of the odd operator $\mathcal{O}$ \cite{Hinschberger2012,Silenko2016PRA}.  

The fields in the last Hamiltonian (\ref{final_FW}) are defined as $\bm{B}=\bm{\nabla}\times\bm{A}$, the external magnetic field, $\bm{E}_{\rm tot} = \bm{E}_{\rm int} + \bm{E}_{\rm ext}$ are the electric fields where $\bm{E}_{\rm int}= -\frac{1}{e}\bm{\nabla}V$ is the internal field that exists even without any perturbation and $\bm{E}_{\rm ext}=-\frac{\partial\bm{A}}{\partial t}$ is the external field (only the temporal part is retained here because of the Coulomb gauge). 

\subsection*{The spin Hamiltonian}

The aim of this work is to formulate the magnetization dynamics on the basis of this Hamiltonian. Thus, we split the Hamiltonian into spin-independent and  spin-dependent parts and consider from now on electrons.
The spin Hamiltonian is straightforwardly given as
\begin{eqnarray}
&\mathcal{H}^{\bm{S}}& (t)  = -\frac{e}{m}\bm{S}\cdot\bm{B} + \frac{e}{4m^3c^2}\left\{\left(\bm{p}-e\bm{A}\right)^2,\bm{S}\cdot \bm{B} \right\}\nonumber\\
 &-&\frac{e}{4m^{2}c^{2}}\bm{S}\cdot\left[ \bm{E}_{\rm tot}\times\left(\bm{p}-e\bm{A}\right)-\left(\bm{p}-e\bm{A}\right)\times\bm{E}_{\rm tot}\right]\nonumber\\
 &-&\frac{ie\hbar}{8m^3c^4}\bm{S}\cdot\left[ \partial_t\bm{E}_{\rm tot}\times\left(\bm{p}-e\bm{A}\right)+\left(\bm{p}-e\bm{A}\right)\times\partial_t\bm{E}_{\rm tot}\right],\nonumber\\
 \label{spin-Ham}
\end{eqnarray}
where the spin operator $\bm{S}=(\hbar/2)\,\bm{\sigma}$ has been used. Let us briefly explain the physical meaning behind each term that appears in $\mathcal{H}^{\bm{S}}(t)$. The first term defines the Zeeman coupling of the electron's spin with the externally applied magnetic field. The second term defines an indirect coupling of light to the Zeeman interaction of spin and the optical $\bm{B}$-field, which can be be shown to have the form of a relativistic Zeeman-like term. The third term implies a general form of the spin-orbit coupling that is gauge invariant \cite{Mondal2015b}, and it includes the effect of the electric field from an internal as well as an external field. The last term is the new  term of relevance here that has  only been considered once in the literature  by Hinschberger {\it et al.}\ \cite{Hinschberger2012}.
Note that, although the last term in Eq.\ (\ref{spin-Ham}) contains the total electric field,  only the time-derivative of the external field plays a role here, because the time derivative of internal field is zero as the ionic potential is time independent. In general if one assumes a plane-wave solution of the electric field in Maxwell's equation as $\bm{E} = \bm{E}_0 e^{i\omega t}$, the last term can be written as $\frac {e\hbar \omega}{8m^3c^4}\bm{S}\cdot (\bm{E}\times\bm{p}) $ and thus adopts the form of a higher-order spin-orbit coupling for a general $E$-field.

The spin-dependent part can be easily rewritten in a shorter format using the identities:  
\begin{align}
\mathcal{A}\times\left(\bm{p}-e\bm{A}\right) - \left(\bm{p}-e\bm{A}\right)\times\mathcal{A} &= 2\mathcal{A}\times \left(\bm{p}-e\bm{A}\right)\nonumber\\ & +i\hbar\bm{\nabla}\times\mathcal{A}\\
\mathcal{A}\times\left(\bm{p}-e\bm{A}\right) + \left(\bm{p}-e\bm{A}\right)\times\mathcal{A} &= - i\hbar\bm{\nabla}\times\mathcal{A}
\end{align}
for any operator $\mathcal{A}$. This allows us to write the spin Hamiltonian as 
\begin{eqnarray}
\label{spin_hamiltonian}
\mathcal{H}^{\bm{S}} &=& -\frac{e}{m}\bm{S}\cdot\bm{B} +\frac{e}{2m^3c^2}\bm{S}\cdot\bm{B}\left[p^2 -2e\bm{A}\cdot \bm{p} +\frac{3e^2}{2}A^2\right]\nonumber\\
& - &\frac{e}{2m^{2}c^{2}}\bm{S}\cdot\big[\bm{E}_{\rm tot}\times\left(\bm{p} -e\bm{A}\right) \big]+\frac{ie\hbar}{4m^2c^2}\bm{S}\cdot\partial_t\bm{B} \nonumber\\
 &+&  \frac{e\hbar^2}{8m^3c^4}\bm{S}\cdot\partial_{tt}\bm{B}\,. 
\end{eqnarray}
Here, the Maxwell's equations have been used to derive the final form that the spatial derivative of the electric field will generate a time derivative of a magnetic field such that $\bm{\nabla}\times\bm{E}_{\rm ext} = -\frac{\partial\bm{B}}{\partial t}$, whilst the curl of a internal field results in zero as the curl of a gradient function is always zero. 
 The final spin Hamiltonian  (\ref{spin_hamiltonian}) bears much importance for the strong laser field-matter interaction as it takes into account all the field-spin coupling terms. It is thus the appropriate fundamental Hamiltonian to understand the effects of those interactions on the magnetization dynamics described in the next section.

\section{Magnetization dynamics}
\label{2}

In general, magnetization is given by the magnetic moment per unit volume in a magnetic solid. The magnetic moment is given by $g\mu_B\langle\bm{S}\rangle$, where $g$ is the Land\'e g-factor and $\mu_B$ is the  unit of Bohr magneton. The magnetization is then written 
\begin{eqnarray}
\label{magnetization}
\bm{M} (\bm{r},t) = \sum_j\frac{g\mu_B}{\Omega} \langle\bm{S}^j\rangle ,
\end{eqnarray}
where $\Omega$ is the suitably chosen volume element, the sum $j$ goes over all electrons in the volume element, and $\langle..\rangle$ is the expectation value.
To derive the dynamics, we take the time derivative in both the sides of Eq.\ (\ref{magnetization}) and, within the adiabatic approximation, we arrive at the equation of motion for the magnetization as \cite{ho04,white2007,Hickey2009}
\begin{eqnarray}
\frac{\partial\bm{M}}{\partial t} & = & \sum_{j}\frac{g\mu_B}{\Omega} \frac{1}{i\hbar}\langle \left[\bm{S}^j,\mathcal{H}^{\bm{S}}(t)\right]\rangle .
\end{eqnarray}
Now the task looks simple, one needs to substitute the spin Hamiltonian (\ref{spin_hamiltonian}) and calculate the commutators in order to find the equation of motion. Note that the dynamics only considers the local dynamics as we have not taken into account the time derivative of particle density operator (for details, see \cite{Mondal2016}). Incorporating the latter would give rise the local as well as non-local processes (i.e., spin currents) within the same footing. 

The first term in the spin Hamiltonian produces the dynamics as
\begin{eqnarray}
\label{1st_dynamics}
\frac{\partial\bm{M}^{(1)}}{\partial t} &=& -\gamma\,\bm{M}\times\bm{B} ,
\end{eqnarray}
with $\gamma =g \vert e \vert /2m $ defines the gyromagnetic ratio and the Land\'e g-factor $g\approx 2$ for spins, the electronic charge $e < 0$. Using the linear relationship of magnetization with the magnetic field $\bm{B} = \mu_0(\bm{H} + \bm{M})$, the latter is replaced in Eq.\ (\ref{1st_dynamics}) to get the usual form in the Landau-Lifshitz equations, $-\gamma_0 \, \bm{M}\times\bm{H}$, where $\gamma_0 = \mu_0\gamma$ is the effective gyromagnetic ratio. This gives the Larmor precession of magnetization around an effective field $\bm{H}$. The effective field will always have a contribution from a exchange field and the relativistic corrections to it, which has not been explicitly taken into account in this article, as they are not in the focus here. For detailed calculations yet including the exchange field see Ref.\ \cite{Mondal2016}.

The second term in the spin Hamiltonian Eq.\ (\ref{spin_hamiltonian}) will result in a relativistic correction to the magnetization precession. Within an uniform field approximation $(\bm{A} = \bm{B}\times\bm{r}/2)$, the corresponding dynamics will take the form
\begin{align}
	\frac{\partial\bm{M}^{(2)}}{\partial t} & = \frac{\gamma}{2m^2c^2}\bm{M}\times\bm{B}\Big\langle p^2 -e\bm{B}\cdot \bm{L} +\frac{3e^2}{8}\left(\bm{B}\times\bm{r}\right)^2\Big\rangle\,,
\end{align} 
with $\bm{L} = \bm{r}\times\bm{p}$ the orbital angular momentum.
The presence of $\gamma/2m^2c^2$ implies that the contribution of this dynamics to the precession is relatively small, while the leading precession dynamics is given by Eq.\ (\ref{1st_dynamics}).
For sake of completeness we note that a relativistic correction to the precession term of similar order $1/m^2c^2$ was obtained previously for the exchange field \cite{Mondal2016}.

The next term in the Hamiltonian is a bit tricky to handle as the third term in Eq.\ (\ref{spin_hamiltonian}) is not hermitian, not even the fourth term which is anti-hermitian. However \textit{together} they form a hermitian Hamiltonian \cite{Mondal2016,widom2009,Hickey_reply2009}. Therefore one has to work together with those terms and cannot only perform the dynamics with an individual term. In an earlier work \cite{Mondal2016} we have shown that taking an uniform magnetic field along with the gauge $\bm{A} = \frac{\bm{B}\times\bm{r}}{2}$ will preserve the hermiticity. The essence of the uniform field lies in the assumption that the skin depth of the electromagnetic field is longer than the thickness of the thin-film samples used in experiments. The dynamical equation of spin motion with the second and third terms thus thus be written in a compact form for harmonic applied fields as \cite{Mondal2016}
\begin{eqnarray}
\label{damping}
\frac{\partial\bm{M}^{(3,4)}}{\partial t} & = & \bm{M}\times\left(A \cdot\frac{\partial\bm{M}}{\partial t} \right),
\end{eqnarray}
with the intrinsic Gilbert damping parameter $A$ that is a tensor defined by
\begin{eqnarray}
A_{ij} & = & \frac{\gamma\mu_0}{4m c^2}\sum_{n,k}\Big[ \langle r_ip_k + p_kr_i \rangle - \langle r_np_n + p_nr_n\rangle\delta _{ik}\Big]\nonumber\\
&& \qquad\qquad\qquad\qquad\qquad \times \Big(\mathbbm{1} + \chi_m^{-1}\Big)_{kj}.
\end{eqnarray}
Here $\chi_m$ is the magnetic susceptibility tensor of rank 2  (a $3\times 3$ matrix) and $\mathbbm{1}$ is the $3 \times 3$ unit matrix. 
Note that for  diagonal terms i.e., $i=k$ the contributions from the expectation values of $r_kp_i$ 
cancel each other. The damping tensor can be decomposed to have contributions from an isotropic Heisenberg-like, anisotropic Ising-like and Dzyaloshinskii-Moriya-like tensors. The anti-symmetric Dzyaloshinskii-Moriya contribution has been shown to lead to a chiral damping of the form $\bm{M} \times ( \bm{D} \times \partial \bm{M} / \partial t )$
\cite{Mondal2016}. 
Experimental observations of chiral damping have been reported recently \cite{Jue2016}.
The other cross term having the form $\bm{E}\times\bm{A}$ in Eq.\ (\ref{spin_hamiltonian}) is related to the angular momentum of the electromagnetic field and thus provides a torque on the spin that has been at the heart of angular magneto-electric coupling \cite{Mondal2015b}. A possible effect in spin dynamics including the light's angular momentum has been investigated in the strong field regime and it has been shown that one has to include this cross term in the dynamics in order to explain the qualitative and quantitative strong field dynamics \cite{bauke14}.

For the last term in the spin Hamiltonian (\ref{spin_hamiltonian}) it is rather easy to formulate the spin dynamics because it is evidently hermitian. Working out the commutator with the spins gives a contribution to the dynamics as
\begin{eqnarray}
\label{2nd_order_time_derivative}
\frac{\partial\bm{M}^{(5)}}{\partial t} & = & \delta\, \bm{M}\times\frac{\partial^2\bm{B}}{\partial t^2}, 
\end{eqnarray}
with the constant $\delta = \frac{\gamma\hbar^2}{8m^2c^4}$.

Let us work explicitly with the second-order time derivative of the magnetic induction by the relation $\bm{B} = \mu_0 (\bm{H} + \bm{M})$, using a chain rule for the derivative:
\begin{eqnarray}
\frac{\partial^2\bm{B}}{\partial t^2}\, && = \frac{\partial}{\partial t} \Big(\frac{\partial \bm{B}}{\partial t}\Big)  = \mu_0\frac{\partial}{\partial t} \Big(\frac{\partial\bm{H}}{\partial t} +\frac{\partial \bm{M}}{\partial t}\Big) \nonumber\\
&& = \mu_0 \Big(\frac{\partial ^2 \bm{H}}{\partial t^2} + \frac{\partial ^2 \bm{M}}{\partial t^2} \Big) .
\end{eqnarray}
This is a generalized equation for the time-derivative of the magnetic induction which can be used even for non-harmonic fields.
The magnetization dynamics is then given by 
\begin{eqnarray}
\frac{\partial\bm{M}^{(5)}}{\partial t} & = & \mu_0 \delta\,\, \bm{M}\times \Big(\frac{\partial ^2 \bm{H}}{\partial t^2} + \frac{\partial ^2 \bm{M}}{\partial t^2} \Big) . 
\end{eqnarray}
Thus the extended LLG equation of motion will have these two additional terms: (1) a field-derivative torque and (2) magnetization-derivative torque, and they appear with their 2$^{\rm nd}$ order time derivative. It deserves to be noted that, in a previous theory we also obtained a similar term--a field-derivative torque in 1$^{\rm st}$ order-time derivative appearing in the generalized Gilbert damping.
Specifically, the extended LLG equation for a general time-dependent field $\bm{H}(t)$ becomes
\begin{eqnarray}
\frac{\partial \bm{M}}{\partial t} & =& - \gamma_0  \bm{M} \times \bm{H}  +   \bm{M}\times \Big[ \bar{A} \cdot \Big( \frac{\partial \bm{H}}{\partial t} + \frac{\partial \bm{M}}{\partial t}\Big)  \Big]
\nonumber \\
&+& \mu_0 \delta\,\, \bm{M}\times \Big(\frac{\partial ^2 \bm{H}}{\partial t^2} + \frac{\partial ^2 \bm{M}}{\partial t^2} \Big) ,
\end{eqnarray}
where $\bar{A}$ is a modified Gilbert damping tensor (for details, see \cite{Mondal2016}). 

However for harmonic fields, the response of the ferromagnetic materials is measured through the differential susceptibility, $\chi_m= \partial\bm{M}/\partial\bm{H}$, because there exists a net magnetization even in the absence of any applied field. With this, the time derivative of the harmonic magnetic field can be further written as: 
\begin{eqnarray}
\frac{\partial ^2 \bm{H}}{\partial t^2} & = & \frac{\partial}{\partial t}\Big(\frac{\partial \bm{H}}{\partial \bm{M}} \frac{\partial \bm{M}}{\partial t}\Big)
 =  \frac{\partial}{\partial t}\Big( \chi_m^{-1} \cdot \frac{\partial \bm{M}}{\partial t} \Big)\nonumber\\
& = & \frac{\partial \chi_m^{-1} }{\partial t} \cdot \frac{\partial \bm{M}}{\partial t} + \chi_m^{-1} \cdot \frac{\partial^2 \bm{M}}{\partial t^2}.
\end{eqnarray}
In general the magnetic susceptibility is a spin-spin response function that is wave-vector and frequency dependent.
Thus, Eq.\ (\ref{2nd_order_time_derivative}) assumes the form with the first and  second order time derivatives as
\begin{eqnarray}
\frac{\partial\bm{M}^{(5)}}{\partial t} & = &  \bm{M}\times\left(\mathcal{K} \cdot\frac{\partial \bm{M}}{\partial t} +   \mathcal{I} \cdot\frac{\partial^2\bm{M}}{\partial t^2}\right),
\end{eqnarray}
where the parameters $\mathcal{I}_{ij} = \mu_0\delta \left(\mathbbm{1} + \chi_m^{-1}\right)_{ij}$ and $\mathcal{K}_{ij} = \mu_0\delta\, \partial _t (\chi^{-1}_m)_{ij}$ are tensors. The  dynamics of the second term is that of the magnetic inertia that operates on shorter time scales \cite{Ciornei2010thesis}. 

Having all the required dynamical terms, finally  the full magnetization dynamics can be written by joining together all the individual parts. Thus the full magnetization dynamics becomes, for harmonic fields, 
\begin{eqnarray}
\frac{\partial\bm{M}}{\partial t} & = & \bm{M}\times\left(-\gamma_0\bm{H} + \Gamma \cdot\frac{\partial\bm{M}}{\partial t} + \mathcal{I}\cdot\frac{\partial^2\bm{M}}{\partial t^2} \right) .
\label{final_dynamics}
\end{eqnarray}
Note that the Gilbert damping parameter $\Gamma$ has two contributions, one from the susceptibility itself, $A_{ij}$, which is of order $1/c^2$ and an other from the time derivative of it, $\mathcal{K}_{ij}$ of order $1/c^4$. Thus, $\Gamma_{ij} =A_{ij}+\mathcal{K}_{ij}$. However we will focus on the first one only as it will obviously be the dominant contribution, i.e., $\Gamma_{ij} \approx A_{ij}$. Even though we consider only the Gilbert damping term of order $1/c^2$ in the discussions, we shall explicitly analyze the other term of the order $1/c^4$. For an ac susceptibility i.e., $\chi_m^{-1} \propto e^{i\omega t} $ we find that $\mathcal{K}_{ij} \propto \mu_0 \delta \,\partial_t (\chi _m ^{-1})_{ij} \propto  i \mu_0\omega \delta\,\chi _m ^{-1}$, which suggests again that the Gilbert damping parameter of the order $1/c^4$ will be given by the imaginary part of the susceptibility, $\mathcal{K}_{ij} \propto - \mu_0\omega \delta\,\mathtt{Im}\big(\chi _m ^{-1}\big)$.  

The last equation (\ref{final_dynamics}) is the central result of this work, as it establishes a rigorous expression for the intrinsic magnetic inertia.
Magnetization dynamics including inertia has been discussed in few earlier articles \cite{Bhattacharjee2012,Henk2012,Ciornei2011,Fahnle2011}. The very last term in Eq.\ (\ref{final_dynamics}) has been associated previously with the inertia magnetization dynamics \cite{Li2015,Olive2012,Tatara2015}. As mentioned, it implies a magnetization nutation i.e., a changing of the precession angle as time progresses. Without the inertia term we obtain the well-known LLG equation of motion that has already been used extensively in magnetization dynamics simulations (see, e.g., \cite{Djordjevic2007,Nowak2007,Skubic2008,evans14,Hinzke2015}).

\section{Discussions}

Magnetic inertia was discussed first in relation to the earth's magnetism \cite{Walker1917}. From a dimensional analysis, the magnetic inertia of a uniformly magnetized sphere undergoing uniform acceleration was estimated to be of the order of $1/c^2$ \cite{Walker1917}, which is consistent with the here-obtained relativistic nature of magnetic inertia.

Our derivation based on the fundamental Dirac-Kohn-Sham Hamiltonian provides explicit expressions for both the Gilbert and inertial dampings. Thus, a comparison can be made between the Gilbert damping parameter and the magnetic inertia parameter of a pure system. As noticed above, both the parameters are given by the magnetization susceptibility tensor, however it should be noted that the quantiy $\langle r_\alpha p_\beta\rangle$ is imaginary itself, because \cite{Mondal2016},
\begin{eqnarray}
\langle r_\alpha p_\beta \rangle &=& -\frac{i\hbar}{2m}\sum_{n,n^\prime,\bm{k}}\frac{f(E_{n\bm{k}}) - f(E_{n^\prime\bm{k}})}{E_{n\bm{k}} - E_{n^\prime\bm{k}}} p^{\alpha}_{nn^\prime} p^{\beta}_{n^\prime n} .\,\,
\end{eqnarray}
Thus the Gilbert damping parameter should be given by the imaginary part of the susceptibility tensor \cite{Gilmore2007thesis,Hickey2009}. On the other hand the magnetic inertia tensor must be given by the real part of the susceptibility \cite{Bhattacharjee2012}. This is in agreement with a recent article where the authors also found the same dependence of real and imaginary parts of susceptibility to the nutation and Gilbert damping  respectively \cite{thonig2016}. In our calculation, the Gilbert damping and inertia parameters adopt the following forms respectively,
\begin{align}
\Gamma_{ij} &= \frac{i\gamma\mu_0}{4m c^2}\sum_{n,k}\left[\langle r_ip_k + p_kr_i \rangle - \langle r_np_n + p_nr_n\rangle\delta _{ik}\right] \nonumber\\
& \qquad\qquad\qquad\qquad\qquad\qquad\qquad\times \mathtt{Im}\big(\chi_m^{-1}\big)_{kj}\nonumber\\ 
& = -\frac{\mu_0\gamma\hbar}{4m c^2}\sum_{n,k}\left[\frac{\langle r_ip_k + p_kr_i \rangle - \langle r_np_n + p_nr_n\rangle\delta _{ik}}{i\hbar}\right]\nonumber\\
& \qquad\qquad\qquad\qquad\qquad\qquad\qquad \times \mathtt{Im}\big(\chi_m^{-1}\big)_{kj}\nonumber\\
& = - \zeta \sum_{n,k}\left[\frac{\langle r_ip_k + p_kr_i \rangle - \langle r_np_n + p_nr_n\rangle\delta _{ik}}{i\hbar}\right]\nonumber\\
& \qquad\qquad\qquad\qquad\qquad\qquad\qquad \times \mathtt{Im}\big(\chi_m^{-1}\big)_{kj}  ,
\end{align}
\begin{align}
\mathcal{I}_{ij} &= \frac{\mu_0 \gamma \hbar^2 }{8m^2c^4}\Big[\mathbbm{1} +\mathtt{Re}\big(\chi_m^{-1}\big)_{kj} \Big]\nonumber\\
&= \frac{\zeta \hbar }{2mc^2}\Big[\mathbbm{1} +\mathtt{Re}\big(\chi_m^{-1}\big)_{kj} \Big] ,
\end{align}
with $\zeta \equiv \frac{\mu_0\gamma\hbar}{4m c^2}  $. 
Note that the change of sign from damping tensor to the inertia tensor that is also consistent with Ref.\ \cite{thonig2016}, and also a factor of 2 present in inertia. However, most importantly, the inertia tensor is $\hbar/mc^2$ times \textit{smaller} than the damping tensor as is revealed in our calculations.
Considering atomic units we can evaluate 
\begin{align*}
\zeta & \sim \frac{\mu_0}{4c^2} \sim \frac{0.00066}{4\times 137^2} \sim 8.8\times10^{-9},\\
\frac{\zeta\hbar }{2mc^2} & \sim \frac{\zeta}{2 c^2} \sim \frac{8.8\times 10^{-9}}{2\times137^2 } \sim 2.34\times 10^{-13} .
\end{align*}
This implies that the intrinsic inertial damping is typically $4 \times 10^4$ times smaller than the Gilbert damping and it is not an independently variable parameter.
Also, because of its smallness  magnetic inertial dynamics will be more significant on shorter timescales \cite{Ciornei2011}.

A further analysis of the two parameters can be made. One can use the Kramers-Kronig transformation to  relate the real and imaginary parts of a susceptibility tensor with one another. This suggests a  relation between the two parameters that has been found by F\"ahnle {\it et al.}\ \cite{fahnle2013PRB}, namely  $\mathcal{I} = - \Gamma \tau $, where $\tau$ is a relaxation time. We obtain here a similar relation, $\mathcal{I} \propto - \Gamma \bar{\tau} $, where $\bar{\tau}=\hbar/mc^2$ has time dimension.

Even though the Gilbert damping is $c^2$ times larger than the inertial damping, the relative strength of the two parameters also depends on the real and imaginary parts of the susceptibility tensor.  
In special cases, when the real part of the susceptibility is much higher than the imaginary part, their strength could be comparable to each other. We note in this context that there exist materials where the real part of the susceptibility is $10^2 - 10^3$ times larger than the imaginary part.  

Finally, we emphasize that our derivation provides the \textit{intrinsic} inertial damping of a pure, isolated system. For the Gilbert damping it is already well known that environmental effects, such as interfaces or grain boundaries, impurities, film thickness, and even interactions of the spins with quasi-particles, for example, phonons, can modify the extrinsic damping (see, e.g., \cite{Kuanr2005,Walowski2008,Song2013}). Similarly, it can be expected that the inertial damping will become modified through environmental influences. An example of environmental effects that can lead to magnetic inertia have been considered previously, for the case of a local spin moment surrounded by conduction electrons, whose spins couple to the local spin moment and affect its dynamics \cite{Bhattacharjee2012,Tatara2015}.

\section{Conclusions}

In conclusion, we have rigorously derived the magnetization dynamics from the fundamental Dirac Hamiltonian and have provided a solid theoretical framework for, and established the origin of, magnetic inertia in pure systems.
We have derived expressions for the Gilbert damping and the magnetic inertial damping on the same footing and have shown that both of them have a relativistic origin. The Gilbert damping stems from a generalized spin-orbit interaction involving external fields, while the inertial damping is due to higher-order (in $1/c^2$) spin-orbit contributions in the external fields. 
Both have been shown to be tensorial quantities. For general time dependent external fields, a field-derivative torque with a 1$^{\rm st}$ order time derivative appears in the Gilbert-type damping, and a  2$^{\rm nd}$ order time-derivative field torque appears in the inertial damping.

In the case of harmonic external fields, the expressions of the magnetic inertia and the Gilbert damping scale with the real part and the imaginary part, respectively, of the magnetic susceptibility tensor, and they are opposite in sign.
Alike the Gilbert damping, the magnetic inertia tensor is also temperature dependent through the magnetic response function and also magnetic moment dependent. 
Importantly, we find that the intrinsic inertial damping is  much smaller than the Gilbert damping, 
which corroborates the fact that magnetic inertia was neglected in the early work on magnetization dynamics \cite{landau35,gilbert56,gilbert04,baryakhtar84}.
This suggests, too, that the influence of magnetic inertia will be quite restricted, unless the real part of the susceptibility is much larger than the imaginary part.
Another possibility to enhance the magnetic inertia would be to use environmental influences to increase its extrinsic contribution.
Our theory based on the Dirac Hamiltonian leads to exact expressions for both the  intrinsic Gilbert and inertial damping terms, thus providing a solid base for their evaluation within \textit{ab initio} electronic structure calculations and giving suitable values that can be used in future LLG magnetization dynamics simulations.

\begin{acknowledgments}

We thank Danny Thonig and Alex Aperis for useful discussions.
This work has been supported by 
the Swedish Research Council (VR), the Knut and Alice Wallenberg Foundation (Contract No.\ 2015.0060), and the Swedish National  Infrastructure for Computing (SNIC).	 

\end{acknowledgments}

\bibliographystyle{apsrev4-1}
%

\end{document}